\documentclass[11pt]{article}
\usepackage[top=0.8in, bottom=1.0in, left=0.7in, right=0.875in]{geometry}
\usepackage{hyperref}
\usepackage{amssymb}
\usepackage{amsmath}
\numberwithin{equation}{section}
\usepackage{slashed}
 \usepackage{blindtext}
\usepackage{amsfonts,amsmath,amssymb,dsfont}
\usepackage{nccmath}
\usepackage{enumerate}
\usepackage{bbm}
\usepackage{nicefrac}
\usepackage[all]{xy}
\usepackage{graphicx}
\usepackage{bbm}
\usepackage{makecell}
\usepackage[table]{xcolor}
\usepackage{color}
\usepackage{longtable}
\usepackage{subcaption}
\usepackage{tikz}
\usepackage{bbold}
\usepackage{multicol}
\usepackage[normalem]{ulem}

\usepackage{upgreek}

\usepackage{float}			% Helps position tables right here [H]
\usepackage{lscape}                 % creates pages in landscape format

\begin{document}
%\begin{titlepage}

 \hfill IFT-UAM/CSIC-23-42
\vspace{15pt}

   \begin{center}
   \baselineskip=16pt

   \begin{Large}\textbf{
\hspace{-18pt} A Positive Energy Theorem for AdS Solitons  \\[8pt]
}
   \end{Large}

\vspace{25pt}

{\large Andrés Anabalon$^{1}$, Mattia Ces\`aro$^{2}$, Antonio Gallerati$^{3}$,  Alfredo Giambrone$^{3,4}$ \,and\\  Mario Trigiante$^{3,4}$}
		
\vspace{25pt}

	\begin{small}
 {\it $^{1}$ Departamento de Ciencias, Facultad de artes Liberales, \\
   Universidad Adolfo Ibáñez, Avenida Padre Hurtado 750, Viña del Mar, Chile}  \\
   
   \vspace{7pt}

	{\it $^{2}$ Departamento de F\'\i sica Te\'orica and Instituto de F\'\i sica Te\'orica UAM/CSIC , \\
   Universidad Aut\'onoma de Madrid, Cantoblanco, 28049 Madrid, Spain}  \\

	\vspace{5pt}
	
	{\it $^{3}$ Politecnico di Torino, Dipartimento di Scienza Applicata e Tecnologia,\\ Corso Duca degli Abruzzi 24, 10129 Torino, Italy}     \\	
 \vspace{5pt}
 {\it $^{4}$ Istituto Nazionale di Fisica Nucleare, Sezione di Torino,\\Via Pietro Giuria 1, 10125 Torino, Italy}     \\	
		
	\end{small}

\vskip 25pt

\end{center}

\begin{center}
\textbf{Abstract}
\end{center}

\begin{quote}
The uncharged AdS$_4$ soliton has been recently shown to be continuously connected to a magnetic, supersymmetric AdS$_4$ soliton within $\mathcal{N}=8$ gauged supergravity. By constructing the asymptotic superalgebra, we establish a positive energy theorem for the magnetic AdS$_4$ solitons admitting well-defined asymptotic Killing spinors, antiperiodic on a contractible $S^1$. We show that there exists only one discrete solution endowed with these boundary conditions satisfying the bound, the latter being saturated by the null energy supersymmetric configuration. Despite having negative energy, the uncharged AdS$_4$ soliton does not contradict the positive energy theorem, as it does not admit well-defined asymptotic Killing spinors.

\end{quote}
\vskip 50pt
%\vfill

%\end{titlepage}
%\tableofcontents

\begin{multicols}{2}

\section{Introduction and Discussion}

%\rc{The spinorial proof of the positive energy theorem in general relativity requires the existence of a spacelike surface where, when restricted to that surface, a spinor satisfies the Dirac equation.}
The spinorial proof of the positive energy theorem in general relativity requires the existence of a spacelike surface and of a spinor field which, when restricted to that surface, satisfies the Dirac equation. These spinors are called Witten spinors  \cite{Witten:1981mf}. Therefore, in contrast, one can ensure that negative energy should imply that such spacelike surface does not exist, namely, there is a non-spacelike singularity and the spacetime is not regular. The latter statement, and in general the proof of the positive energy theorem, %\rc{This statement}
has been extended to asymptotically AdS spaces and gauged supergravity in several papers \cite{Gibbons:1983aq,Breitenlohner:1982bm,Breitenlohner:1982jf}. The AdS soliton seems to be at odds with this expectation \cite{Horowitz:1998ha}. Indeed, it is an everywhere regular solution with negative total energy. However, spinors must be antiperiodic around a circle $S^{1}$ that smoothly %\rc{closes} 
contracts to a point in the interior of the AdS soliton spacetime. This boundary condition excludes the existence of asymptotic Killing spinors and of Witten spinors, seemingly breaking all supersymmetry. Indeed, the antiperiodicity of the spinors on an $S^1$ is typically called a supersymmetry breaking boundary condition, see for instance \cite{Horowitz:1998ha}. As such, it is expected to preclude the possibility of constructing a positive energy theorem whenever there is an $S^1$ that smoothly contracts to a point in the bulk of the spacetime. Indeed, as can be seen from the abstract of \cite{Horowitz:1998ha} solutions with a spacelike cycle $S^1$ at the boundary are known as supersymmetry breaking boundary conditions.  This issue has been discussed in the literature \cite{Cheng:2005wk}, but until now the problem of endowing these solutions with a positive energy theorem has remained open.

We have recently shown that by including a magnetic flux, the same boundary conditions on the metric that yield the AdS soliton of \cite{Horowitz:1998ha}, yield a supersymmetric soliton in the gauged $\mathcal{N}=8$ supergravity \cite{Anabalon:2021tua,Anabalon:2022aig}. This supersymmetric solution has zero energy. In this letter, we prove that the existence of this supersymmetric state should imply that all the states with the right boundary conditions should have an energy that is larger than the energy of the supersymmetric state. Hence, establishing a positive energy theorem for the case when the spinors are antiperiodic on a contractible $S^1$. This is done by inspecting the supersymmetry algebra on the soliton solution and deriving from the anticommutator of two supersymmetries a BPS bound. This amounts to applying  the approach of \cite{Hristov:2011ye, Hristov:2011qr}, in the presence Wilson lines, to this new class of solutions.

There is an apparent tension between the two paragraphs above. Indeed, we are stating that it is possible that an everywhere regular negative-energy solution is continuously connected with a supersymmetric solution, which in turn ensures that the energy must be positive. We will show below that this issue is resolved in gauged supergravity because our positive energy theorem only constrains solutions that support asymptotic Killing spinors with antiperiodic boundary conditions. It turns out that the solutions with these boundary conditions are not continuously connected and form a discrete set within the family we study here. Moreover, all the negative energy solutions do not have antiperiodic asymptotic Killing spinors and there is only one non-supersymmetric solution with positive energy supporting them. A similar issue was discussed in \cite{Anabalon:2022aig} in relation to supersymmetric hairy solutions. 

What is remarkable about this result is that the Einstein-Maxwell theory, with a cosmological constant, is not only a universal sector of a number of maximal gauged supergravity theories, but, as part of a pure $\mathcal{N}=2,\,D=4$ supergravity with a cosmological constant, it is a consistent truncation of $D=11$ supergravity \cite{Duff:1985jd}. More generally, pure $\mathcal{N}$-extended four-dimensional supergravity was conjectured in \cite{Gauntlett:2007ma} and proven in \cite{Cassani:2019vcl}, to provide a consistent truncation of Type II  or $D=11$ supergravities on a generic background with an ${\rm AdS}_4$ factor, preserving $\mathcal{N}$ four-dimensional supersymmetries. This implies that Einstein-Maxwell theory with a cosmological constant, as a consistent truncation of a pure $\mathcal{N}=2,\,D=4$ supergravity, is a consistent truncation of a generic background, solution to   Type II  or $D=11$ supergravities, with an ${\rm AdS}_4$ factor and preserving $\mathcal{N}=2$ supersymmetries.  In light of this, the results provided in this paper ensure the non-perturbative existence of well-defined ground states in a large class of quantum field theories at strong t'Hooft coupling, even when there is a compact $S^1$ direction in the spacetime.

Finally, it would be very interesting to extend this study to the case where the solutions are spinning. In asymptotically AdS spacetimes the superalgebra contains the angular momentum in the RHS of the anticommutators of the supercharges. As such, the BPS bound would be modified and it might be possible that there are negative energy solutions satisfying a BPS bound. We leave this question open for future research.

This letter is organized as follows. In  section \ref{sec:sugramodel} we review our conventions and provide some details of the supergravity model. In section \ref{sec:susyalgebrafromvariations} we summarise and report the techniques from \cite{Hristov:2011ye,Hristov:2011qr} in order to read off the asymptotic superalgebra and the related BPS bound of a generic $\mathcal{N}=2$, asymptotic AdS (or magnetic AdS (mAdS)) supergravity configuration. Section \ref{sec:AsymptoticSUSY} is devoted to applying such techniques to the family of magnetic soliton solutions of interest, and to derive their BPS bound for the cases in which the asymptotic Killing spinors exist and are well-defined.
%Section \ref{sec:conclusions} concludes by providing further comments.

%
\section{Gauged $\mathcal{N}=2$ supergravity} \label{sec:sugramodel}

The Einstein-Maxwell-AdS theory furnishes % \bc{describes} 
the bosonic sector of the minimal gauged $\mathcal{N=}2$ supergravity in four dimensions, describing the gravity multiplet in the presence of a cosmological constant. The bulk
action reads \cite{Anabalon:2020pez}
\begin{equation}
S\left(  g,A\right)  =\frac{1}{\kappa}\int d^{4}x\sqrt{-g}\left[  \frac{R}{2}-\frac{1}%
{8}F_{\mu\nu}F^{\mu\nu}+\frac{3}{\ell^{2}}\right] \, , \label{Lag}%
\end{equation}
where $F(A)_{\mu\nu}=\partial_{\mu}A_{\nu}-\partial_{\nu}A_{\mu}\,.$The field
equations are%
\begin{align}
&  \partial_{\mu}\big(\sqrt{-g}F^{\mu\nu}\big)=0\,\,,\nonumber\\[3pt]
&  \nonumber R_{\mu\nu}-\tfrac{1}{2}g_{\mu\nu}R-\tfrac{1}{2}\big[F_{\mu\rho}\,F_{\nu}%
{}^{\rho}-\tfrac{1}{4}g_{\mu\nu}F_{\rho\sigma}F^{\rho\sigma}\big]\\
 & \nonumber\hspace{6cm}-\frac
{3}{\ell^{2}}\,g_{\mu\nu}=0\,.\\
\end{align}
When supplemented by the fermionic sector, the theory is invariant
under supersymmetric transformations of all the fields. We
shall only make use of the transformation of the Rarita-Schwinger fields
$\psi_{\mu}{\!}^{i}\,$:

\begin{align}\label{eq:gravitinovariation}
\nonumber\delta\psi_{\mu}{\!}^{i}=\;  &  2\,\mathcal{D}_{\mu}\epsilon^{i}-\tfrac{1}%
{4}F(A)_{\rho\sigma}\gamma^{\rho\sigma}\gamma_{\mu}\,\varepsilon
^{ij}\,\epsilon_{j}\\[2pt]
&\hspace{1.3cm}+\ell^{-1}\,\varepsilon^{ij}\,t_{j}{}^{k}\,\gamma_{\mu
}\epsilon_{k}\equiv \tilde{\mathcal{D}}_\mu \epsilon^i\,,\\[5pt]
\nonumber\delta\psi_{\mu\,i}=\;  &  2\,\mathcal{D}_{\mu}\epsilon_{i}-\tfrac{1}%
{4}F(A)_{\rho\sigma}\gamma^{\rho\sigma}\gamma_{\mu}\,\varepsilon
_{ij}\,\epsilon^{j}\\[2pt]
&\hspace{1.3cm}+\ell^{-1}\,\varepsilon_{ij}\,t^{j}{}_{k}\,\gamma_{\mu
}\epsilon^{k}\equiv \tilde{\mathcal{D}}_\mu \epsilon_i\,,
\end{align}
where $\gamma^{5}=-\mathrm{i}\gamma^{0}\gamma^{1}\gamma^{2}\gamma^{3}$,
$\gamma^{5}\psi_{\mu}^{i}=\psi_{\mu}^{i}$, $\gamma^{5}\psi_{\mu i}%
=-\psi_{\mu i}$ and $i=1,2$, and we shall pick $t^{i}{}_{j}=$ $\mathrm{i}%
\sigma_{3}\Longrightarrow t_{i}{}^{j}=$ $-\mathrm{i}\sigma_{3}$. The covariant
derivatives of the supersymmetry parameters are given by
\begin{align}
\mathcal{D}_{\mu}\epsilon^{i}=  &  \;\big(\partial_{\mu}+\tfrac{1}{4}%
\omega_{\mu}{\!}^{ab}\gamma_{ab}\big)\epsilon^{i}-\tfrac{1}{2\ell}A_{\mu
}\,t^{i}{\!}_{j}\,\epsilon^{j}\,,\label{eq:covariant-epsilon-der}\\[1mm]
\mathcal{D}_{\mu}\epsilon_{i}=  &  \;\big(\partial_{\mu}+\tfrac{1}{4}%
\omega_{\mu}{\!}^{ab}\gamma_{ab}\big)\epsilon_{i}-\tfrac{1}{2\ell}A_{\mu
}\,t_{i}{}^{j}\,\epsilon_{j}\,.
\end{align}
\subsection{The AdS Soliton with a magnetic flux}

%\rc{It is our interest to consider}
We shall focus, in what follows, on a class
of soliton solutions that can be partially supersymmetric. The solutions of interest are characterized by a space-time metric of the form \cite{Anabalon:2021tua}
\begin{equation}\label{eq:metric}
ds^2=\frac{r^2}{\ell^2}(-dt^2+dz^2)+\frac{dr^2}{f(r)}+f(r)d\phi^2\, ,
\end{equation}
with
\begin{align}
f(r)  &  =\frac{r^{2}}{\ell^{2}}-\frac{\mu}{r}-\frac{Q^{2}}{r^{2}}\,,\label{eq:fr}
\end{align}
and a graviphoton 1-form field
\begin{align}
A  &  =\left(  \frac{2Q}{r}-\frac{2Q}{r_{0}}\right)  d\phi\text{ ,}
\end{align}
where $r_{0}$ is the largest root of the equation $f(r_{0})=0$. The coordinate
$r$ takes its values in the half-interval $\left[  r_{0},\infty\right]  $.
Regularity of the metric requires $\phi\in\left[  0,\Delta\right] $, while the coordinate z can be chosen to be either compact or non-compact.
Regularity of the metric at $r=r_0$ also implies

\begin{equation}
\Delta=\frac{4\pi\ell^{2}r_{0}^{3}}{3r_{0}^{4}+Q^{2}\ell^{2}}\, . \label{HP}
\end{equation}
The net magnetic flux along the $z$ axis is
\begin{equation}
    \Phi=-\int A_{\phi}(r=\infty)d\phi=\frac{2Q}{r_0}\Delta\,.
\end{equation}

The dual energy-momentum tensor reads
\begin{equation}\label{eq:energymomtensor}
\left\langle T_{tt}\right\rangle=-\frac{\mu}{2\kappa \ell^{2}}\text{ ,}%
\quad\left\langle T_{zz}\right\rangle =\frac{\mu}{2\kappa \ell^{2}}\text{
,}\quad\left\langle T_{\phi\phi}\right\rangle =-\frac{\mu}{\kappa \ell^{2}}\, .% 
\end{equation}
Therefore, $\mu$ is proportional to the energy density of the soliton. The gauge field gives a v.e.v. for the current in the boundary theory:
\begin{equation}
\left\langle J^{\nu}\right\rangle =\frac{\delta S}{\delta A_{\nu}}=-\frac{1}{2 \kappa}N_{\mu}F^{\mu\nu}\sqrt{\left\vert h\right\vert }=\frac{Q}{\kappa \ell^2 }\delta_{t_b}^{\nu}\, .
\end{equation}

This soliton was found in \cite{Astorino:2012zm} and studied for fixed gauge field boundary condition in \cite{Kastor:2015wda}, where it was pointed out that for each value of the boundary conditions ($\Delta$, $\Phi$) there are two solutions, one with non-negative energy, continuously connected with AdS$_4$ and the other continuously connected with the AdS soliton. The fixed $J^{\alpha}$ boundary condition is also possible in four dimensions for gauge fields \cite{Marolf:2006nd}. This alternative quantization for this solution was studied in \cite{Anabalon:2021tua, Anabalon:2022aig}. In \cite{Anabalon:2021tua} it was pointed out that the $\mu=0$ solution is supersymmetric. We shall quickly review this next.

\subsection{Supersymmetric AdS Solitons}

The vierbein of the solution can be chosen of the form
\begin{eqnarray}\label{eq:framecomponents}
\nonumber e^0&=&\frac{r}{\ell}dt\, ,\\
\nonumber e^1&=&\frac{dr}{\sqrt{f(r)}}\, ,\\
\nonumber e^2&=&\sqrt{f(r)}d\phi\, ,\\ 
 e^3&=&\frac{r}{\ell}dz\, ,
\end{eqnarray}
whereas the non-vanishing components of the spin-connection are
\begin{equation}\label{eq:spinconnection}
\omega_t{}^{01}=\frac{\sqrt{f(r)}}{\ell}\, , \quad \omega_\phi{}^{21}=\frac{f^\prime}{2}\,,\quad \omega_z{}^{31}=\frac{\sqrt{f(r)}}{\ell}\, ,
\end{equation}
with
\begin{eqnarray}\label{eq:derf}
 f^\prime=\frac{2r}{\ell^2}+\frac{\mu}{r^2}+2\frac{Q^2}{r^3}\,. 
\end{eqnarray}
The basis elements of the Clifford algebra are chosen as follows 
\begin{eqnarray}\label{eq:Cliffordalgebra}
\nonumber
&&\gamma^0=-i\begin{pmatrix}
0&\sigma_2\\
\sigma_2 & 0
\end{pmatrix}\, , \quad 
\gamma^1=-\begin{pmatrix}
\sigma_3 & 0\\
0&\sigma_3
\end{pmatrix}\, ,\\
\nonumber&&\gamma^2=\hspace{3mm}i\begin{pmatrix}
0& -\sigma_2\\
\sigma_2 & 0
\end{pmatrix}\, ,\quad
\gamma^3=\begin{pmatrix}
\sigma_1 & 0\\
0& \sigma_1
\end{pmatrix}\, .\\
\end{eqnarray}
After performing the following change in the radial coordinate on the solution with $\mu=0$
\begin{equation}\label{eq:changeofcoord}
r=r_0\sqrt{\cosh(\rho)}\, ,
\end{equation}
and working with the combinations
\begin{equation}\label{eq:spinorscombo}
\chi_1\equiv \epsilon^1_{(1)}+\epsilon_{2\;(1)}\, ,\qquad \chi_2=\epsilon^1_{(2)}+\epsilon_{2\;(2)}\,,
\end{equation}
the Killing spinors read
\begin{eqnarray}\label{eq:Killingspinors}
\nonumber\chi_1=e^{-i\pi\tfrac{\phi}{\Delta }}\frac{1}{\cosh(\rho)^{1/4}} \begin{pmatrix}
           \sinh{\tfrac{\rho}{2}} \\
          -\cosh{\tfrac{\rho}{2}} \\
          i\cosh{\tfrac{\rho}{2}} \\
           i\sinh{\tfrac{\rho}{2}}
         \end{pmatrix}\, ,\\
\chi_2=e^{-i\pi\tfrac{\phi}{\Delta }}\frac{1}{\cosh(\rho)^{1/4}} \begin{pmatrix}
          -\cosh{\tfrac{\rho}{2}}  \\
          \sinh{\tfrac{\rho}{2}} \\
          i\sinh{\tfrac{\rho}{2}}\\
           i\cosh{\tfrac{\rho}{2}} 
         \end{pmatrix}\, .
\end{eqnarray}
It is useful to label the different spinorss with the index $A=1,2$, together with their SU(2) upper and lower indices $i=1,2$, namely
\begin{equation}
\epsilon^i_{(A)}\, , \qquad \epsilon_{i\;(A)}=(\epsilon^i_{(A)})^*\, .
\end{equation}
From \eqref{eq:Killingspinors} it is easy to recover 
\begin{eqnarray}\label{eq:Weylcomponents}
\nonumber\epsilon^1_{(1)}=\tfrac{1}{2}(\mathbb{1}+\gamma^5)\chi^1,&&\epsilon_{2\;(1)}=\tfrac{1}{2}(\mathbb{1}-\gamma^5)\chi^1\,,\\
\nonumber
\epsilon^1_{(2)}=\tfrac{1}{2}(\mathbb{1}+\gamma^5)\chi^2, &&\epsilon_{2\;(2)}=\tfrac{1}{2}(\mathbb{1}-\gamma^5)\chi^2\, ,\\
\end{eqnarray}
and their complex conjugates
\begin{eqnarray}\label{eq:Weylconjugates}
\nonumber\epsilon_{1\;(1)}=(\epsilon^1_{(1)})^*, \quad\epsilon^{2}_{(1)}=(\epsilon_{2\;(1)})^*\,,\\
\epsilon_{1\;(2)}=(\epsilon^1_{(2)})^*,\quad\epsilon^2_{(2)}=(\epsilon_{2\;(2)})^*\, .
\end{eqnarray}

\subsection{Asymptotic anti-periodic Killing spinors}

It follows from the phase of \eqref{eq:Killingspinors} that the Killing spinors are indeed anti-periodic when moved around the $S^1$ cycle $\phi \rightarrow \phi + \Delta $. As these Killing spinors solve the Killing spinor equation everywhere in the four-manifold, their existence implies that there should be a BPS bound that this configuration saturates. As we shall show below, the BPS bound indeed implies that the energy must be positive. Hence, we would like to discuss now specifically how it happens that the uncharged AdS-soliton or the AdS-soliton with a Wilson line violate this bound.

To understand this, it is useful to consider the Killing spinors of a locally AdS$_4$ spacetime in the presence of a constant $U(1)$ connection, $A=A_{\phi}d\phi$. For this case, we have that the local
solutions to the Killing spinor equation are%
\begin{equation}
\label{eq:KSADS}
\scriptstyle\chi_{1}^{AdS}=\exp\left(  \frac{\mathrm{i}A_{\phi}}{2\ell}\phi\right)
r^{1/2}\scriptstyle\left(
\begin{array}
[c]{c}%
1\\
-1\\
0\\
0
\end{array}
\right)  \text{,}\;\,\chi_{2}^{AdS}=\exp\left(  \frac{\mathrm{i}A_{\phi}%
}{2\ell}\phi\right)  r^{1/2}\left(
\begin{array}
[c]{c}%
0\\
0\\
1\\
1
\end{array}
\right)  \text{,}%
\end{equation}%
\begin{eqnarray}
\nonumber\scriptstyle\chi_{3}^{AdS}=\exp\left(  \frac{\mathrm{i}A_{\phi}}{2\ell}\phi\right)
\left(
\begin{array}
[c]{c}%
r^{1/2}\left(  t+\phi\right) \\
-r^{1/2}\left(  t+\phi\right) \\
-r^{1/2}z-\ell^{2}r^{-1/2}\\
-r^{1/2}z+\ell^{2}r^{-1/2}%
\end{array}
\right)  \text{,} \\
\scriptstyle\chi_{4}^{AdS}=\exp\left(  \frac{\mathrm{i}A_{\phi}%
}{2\ell}\phi\right)  \left(
\begin{array}
[c]{c}%
-r^{1/2}z+\ell^{2}r^{-1/2}\\
r^{1/2}z+\ell^{2}r^{-1/2}\\
r^{1/2}\left(  t-\phi\right) \\
r^{1/2}\left(  t-\phi\right)
\end{array}
\right)  \,.
\end{eqnarray}
The spinors $\chi_{3}^{AdS}$ and $\chi_{4}^{AdS}$, are not
invariant under the identification which makes $\phi$ periodic, which therefore breaks
at least half the supersymmetry. The spinors $\chi_{1}^{AdS}$ and $\chi
_{2}^{AdS}$, are invariant (up to sign) if $A_{\phi}=2\pi n\ell/\Delta$ for integer $n$. 

Now, it is easy to understand why the original AdS-soliton of Horowitz and Myers can not have asymptotic Killing spinors that approach these for large $r$. There is no asymptotically constant $U(1)$ connection which is regular at the axis of symmetry and has vanishing energy-momentum tensor. Hence, an asymptotic Killing spinor on this background can not be anti-periodic and is therefore not in the same class of boundary conditions as those solutions which are 
%\rc{that is} 
constrained by our BPS bound.

A natural question to ask now is  whether there is any non-trivial configuration sharing the same boundary conditions as the supersymmetric AdS soliton and having positive energy, as implied by the corresponding BPS bound. %\rc{with positive energy due to a BPS bound associated with the supersymmetric AdS soliton}.
From the form of the spinors \eqref{eq:KSADS} we see that any solution that asymptotically approaches a local AdS$_4$ spacetime can support an antiperiodic spinor, which asymptotically satisfies the Killing spinor equation, provided the asymptotic form of the connection and the period satisfies $A_{\phi} \Delta=2\pi n\ell$, for odd-integer $n$. For the non-supersymmetric magnetic solitons presented in this paper, it follows that  
\begin{equation}
\frac{-A_{\phi} \Delta}{2 \ell}=\frac{Q }{r_0 \ell}\Delta=\frac{4\pi\ell r_{0}^{2} Q}{3r_{0}^{4}+Q^{2}\ell^{2}}=n \pi\, .
\end{equation}

This equation has real solutions only for $n=1$. The solutions are $r_0=\sqrt{Q \ell}$ and $r_0=\sqrt{\frac{Q \ell}{3}}$. The first case, $r_0=\sqrt{Q \ell}$ is the supersymmetric soliton and the second case is a non-supersymmetric soliton with positive energy and with the same boundary conditions as the BPS soliton. Indeed, the energy density of the non-supersymmetric solution with $r_0=\sqrt{\frac{Q \ell}{3}}$  is 
\begin{equation}
\left\langle T_{tt}\right\rangle=-\frac{\mu}{2 \ell \kappa}=\frac{4 Q^{3/2} \sqrt{3}}{9 \ell^{3/2}}\, .
\end{equation}

Now, let us move on to discussing the derivation %\rc{analyze the construction} 
of the bound from the BPS algebra.
\section{Asymptotic superalgebra from supersymmetry variations} \label{sec:susyalgebrafromvariations}
%Conserved charges in General relativity are defined as integrals of conserved Noether currents along timelike hypersurfaces at the asymptotic infinity. The reason to go all the way to infinity is to ensure that the charge such defined is coordinate-indipendent, and that it preserves at least a Poincar\'e subset of the full diffemorphism invariance group. The same rationale can be followed in $\mathcal{N}=2$ supergravity for the definition of the conserved  supercharges. The Noether super-current built from the supersymmetry variation parameter, the Killing spinor, provides the on-shell conserved supercharge, once integrated at spatial infinity. 
In \cite{Hristov:2011ye} (see \cite{Hristov:2011qr} for an extension to $\mathcal{N}=2$ supergravity coupled to matter) a methodology providing the asymptotic superalgebras and conserved charges of $\mathcal{N}=2$ supergravity solutions that asymptote respectively to AdS or mAdS spacetimes, and their relative BPS bounds, was crafted. Crucially, such method relies on the existence of asymptotic Killing spinors. In this section we turn to briefly recall how this procedure is implemented.

Let $\Psi_\mu{}^i$ be a Majorana gravitino and $\chi^i$, $i=1,2$, be the Majorana Killing spinors of the generic $\mathcal{N}=2$ bulk configuration. Under supersymmetry variation with respect to $\chi^i$, the Noether supercurrent $\mathcal{J}_\chi{}^{\mu}$ is defined, in our conventions, as 
\begin{equation}
 \mathcal{J}_\chi{}^{\mu}=\;i\epsilon^{\mu\nu\rho\sigma}\overline{\Psi}_{\nu\,i}\,\gamma_5\gamma_{\rho}\tilde{\mathcal{D}}_\sigma\chi^i \, ,  
\end{equation}
where $\gamma_\rho$ is the flat gamma matrix $\gamma_a$ contracted with the vielbein $e_\rho{}^a$, and where the operator $\tilde{\mathcal{D}}$ is the one appearing in the gravitino variation under supersymmetry,
\begin{equation}
\delta_\chi \Psi_\mu{}^i=\tilde{\mathcal{D}}_\mu\chi^i\,.
\end{equation} In principle, such supercurrent is defined up to improvement terms, but a comparison between the canonical Dirac brackets and the supersymmetry variations uniquely specifies it (see \cite{Hristov:2011ye} for details).
The (scalar) Noether supercharge is then covariantly defined, in our notations, as
\begin{equation}\label{eq:supercharge1}
\mathcal{Q}=\int_V d\Sigma_\mu\mathcal{J}_\chi{}^{\mu}=-i\int_Vd\Sigma_\mu\epsilon^{\mu\nu\rho\sigma}\,\overline{\Psi}_{\sigma\;i} \gamma_5 \gamma_\rho \tilde{\mathcal{D}}_\nu\chi^i\,, 
\end{equation}
namely as the volume integral over the spacelike hypersurface V. Upon integration by parts and after imposing the field equations, one can rewrite \eqref{eq:supercharge1} as
\begin{equation}\label{eq:supercharge2}
\mathcal{Q}=-i\int_{\partial V}d\Sigma_{\mu\nu}\epsilon^{\mu\nu\rho\sigma}\,\overline{\Psi}_{\sigma\;i} \gamma_5 \gamma_\rho \chi^i\, ,
\end{equation}
where we have defined 
\begin{equation}
d\Sigma_{\mu\nu}=\frac{1}{2}\epsilon_{\mu\nu\rho\sigma} dx^\rho\wedge dx^\sigma\, ,   
\end{equation}
namely as the surface integral along the spacelike hypersurface at radial infinity $\partial V$. As customary in general relativity, the integral should be performed at asymptotic infinity with respect to the radial coordinate: this is the only way to ensure both the coordinate independence and to guarantee at least Poincar\'e invariance of the charge so defined. The anticommutator between two supercharges simply amounts to the supercharge variation itself
\begin{flalign}\label{eq:superchargesanticomm}
\nonumber &\{\mathcal{Q},\mathcal{Q}\}=\delta_\chi \mathcal{Q}=-i\int_{\partial V}d\Sigma_{\mu\nu}\epsilon^{\mu\nu\rho\sigma}\,\left(\delta_\chi\overline{\Psi}_{\sigma\;i} \right)\gamma_5 \gamma_\rho \chi^i\\
&\hspace{2.3cm}=-i\int_{\partial V}d\Sigma_{\mu\nu}\epsilon^{\mu\nu\rho\sigma}\overline{\chi}_i \gamma_5 \gamma_\rho\tilde{\mathcal{D}}_\sigma \chi^i\,\;\;\; .
\end{flalign}
If the right-hand-side of \eqref{eq:superchargesanticomm} is computed on a supersymmetry-preserving background (e.g. the vacuum of the theory) and $\chi^i$ are chosen to be the corresponding Killing spinors, the anticommutator in \eqref{eq:superchargesanticomm} obviously vanishes. If, on the other hand, we compute eq. \eqref{eq:superchargesanticomm} on a (non-necessarily supersymmetric) bulk configuration that is different from the vacuum, but which asymptotes it, the bulk configuration itself can be viewed as an excitation on top of the vacuum state.
In particular, although global Killing spinors on the bulk configuration may not exist, we can define for it  \emph{asymptotic Killing spinors $\chi^{i}{}_{\infty}$} obtained by solving the Killing spinor equations at radial infinity, and which coincide, in this limit, with the Killing spinors of the vacuum configuration. Therefore, the insertion of $\chi^{i}{}_{\infty}$ into \eqref{eq:supercharge2} yields now, through \eqref{eq:superchargesanticomm}, a non-vanishing result, describing the realization of the vacuum superalgebra on the bulk configuration.  It is here that the asymptotic existence and well-definiteness of the Killing spinor $\chi$ is needed.  It is also useful to remark that, while $\chi^{i}{}_{\infty}$ are defined at spatial infinity, the details of the bulk geometry are all still encoded in the operator $\tilde{\mathcal{D}}$ and in the vielbein contracting the flat gamma matrices $\gamma$ in \eqref{eq:superchargesanticomm}.

To make things explicit, after plugging $\chi^{i}{}_{\infty}$ into \eqref{eq:supercharge2}, and having obtained \eqref{eq:superchargesanticomm}, the final step is to extract the spinorial structure out of the scalar supercharges, 
\begin{equation}\label{eq:spinorstructure}
\mathcal{Q}=(\epsilon_0^i)_\alpha \;\mathcal{Q}^\alpha{}_i\, ,
\end{equation}
in terms of a doublet of constant Majorana spinors $\epsilon_0^i$. The asymptotic superalgebra can be then read, schematically, from 
\begin{eqnarray}\label{eq:asymptoticsuperalg}
\nonumber&&(\epsilon_0{}^i)_\alpha\,\{\mathcal{Q}^\alpha{}_i,\mathcal{Q}_{\beta\;j}\}\,(\epsilon_0{}^{\beta \,j})=\\
&&\nonumber\hspace{2.5cm}(\epsilon_0{}^i)_\alpha\;\left(C(T,t)\;T^{\alpha}{}_{\beta}\;t_{ij}\right)\;(\epsilon_0{}^{\beta \;j})\,,\\
\end{eqnarray}
where $T^{\alpha}{}_{\beta}$ and $t_{ij}$ are the relevant superalgebra generators coming in representations of respectively SO$(2,3)$ and SO(2), and $C(T,t)$ the associated coefficients giving rise to the conserved bosonic charges. For generic asymptotic configurations, this SO$(2,3)\times$SO(2) structure will be broken, as in the case of mAdS \cite{Hristov:2011ye}.  Given the asymptotic superalgebra \eqref{eq:asymptoticsuperalg}, standard procedures (see \textit{e.g.} \cite{Gauntlett:2000ch}) allow for the extraction of the related BPS bound.

In the following, we will apply this recipe to the soliton configuration, in order to derive the relevant BPS bound in cases where asymptotic Killing spinors exist and are well-defined. 
\section{Asymptotic superalgebra and BPS bound}\label{sec:AsymptoticSUSY}
\subsection{Asymptotic Killing spinors}
In order to make contact with the notation employed in \cite{Hristov:2011ye}, we find convenient to define the Majorana spinors
\begin{equation}\label{eq:Majorana}
\chi^i=\epsilon^i+ (\epsilon^i)^*\, .
\end{equation}
The action of the derivative operator on such a spinor can be inferred from the structure \eqref{eq:gravitinovariation}, 
\begin{eqnarray}\label{eq:tildeDonlambda}
\nonumber\tilde{\mathcal{D}}_\mu \chi^i&=&2\left(\partial_\mu+\tfrac{1}{4}\omega_{\mu}{}^{ab}\gamma_{ab}\right)\chi^i-\tfrac{i}{\ell}A_\mu\,\gamma^5 (\sigma^3)^i{}_j\, \chi^j\,\\
&&\nonumber-\tfrac{i}{4}F_{\rho\sigma}\gamma^{\rho\sigma}\gamma_\mu (\sigma^2)^i{}_j\chi^j-\frac{i}{\ell}\gamma_\mu \gamma^5 (\sigma^1)^i{}_j \chi^j\,.\\
\end{eqnarray}
The asymptotic Majorana spinors are computed explicitly by solving the asymptotic Killing spinor equations, and their form is 
\begin{eqnarray}\label{eq:Majoranaspinorsexplicitform}
\nonumber\chi^i&=&r^{1/2}\mathbb{P}^i{}_k\mathcal{O}^k{}_j(\phi) \epsilon_0^j\\
\nonumber&=&r^{1/2}\frac{1}{2}\left(\delta^i_k+i \gamma_1\gamma^5 (\sigma^1)^i{}_k\right)\\
\nonumber && \hspace{1.4cm}\times\left(\text{Exp}\left[-i \scriptstyle{\frac{Q}{\ell {r_0}}\phi \gamma^5 (\sigma^3)}\right]\right)^k{}_j \,\epsilon_0^j\, ,\\
\end{eqnarray}
where $\epsilon_0^j$ is a doublet of arbitrary constant Majorana spinors. Alternatively, \eqref{eq:Majoranaspinorsexplicitform} can be rewritten by expanding the definition of the exponential function of operators as
\begin{eqnarray}\label{eq:Majoranaspinorsexplicitformnicer}
\nonumber\chi^i&=&r^{1/2}\frac{1}{2}\left(\delta^i_k+i \gamma_1\gamma^5 (\sigma^1)^i{}_k\right)\\
&&\nonumber\times\left(\cos{\left(\scriptstyle\frac{Q}{\ell {r_0}}\phi\right)}\delta^k_j-i \sin{\left(\scriptstyle\frac{Q}{\ell {r_0}}\phi\right)} \gamma^5 (\sigma^3)^k{}_j\right)\epsilon_0^j\, .\\
\end{eqnarray}
Let us prove that this is the solution to the asymptotic Killing spinor equations. The Killing spinor equations in the bulk read
\begin{flalign}
&\nonumber\medmath{2\partial_t \chi^i+\frac{1}{2}\omega_t^{ab}\gamma_{ab}\chi^i+\frac{Q}{r^2}\gamma_{31}\gamma_t(i\sigma^2)^i{}_j\chi^j-i\frac{\gamma_t}{\ell}\gamma^5(\sigma^1)^i{}_j\chi^j=0}\, ,\label{eq:Killingspinoreqtst1} \\
\\[4pt]
&\nonumber\medmath{2\partial_r \chi^i+\frac{Q}{r^2}\gamma_{31}\gamma_r(i\sigma^2)^i{}_j\chi^j-i\frac{\gamma_r}{\ell}\gamma^5(\sigma^1)^i{}_j\chi^j=0}\, \\\label{eq:Killingspinoreqtsr1}\\[4pt]
&\nonumber\medmath{2\partial_\phi \chi^i+\frac{f^\prime}{2}\gamma_{21}\chi^i-i\frac{2Q}{\ell}\left(\frac{1}{r}-\frac{1}{r_0}\right)\gamma^5(\sigma^3)^i{}_j\chi^j}\\
&\hspace{1.2cm}\medmath{+\frac{Q}{r^2}\gamma_{31}\sqrt{f}\gamma_2(i\sigma^2)^i{}_j\chi^j-i\frac{\sqrt{f}}{\ell}\gamma_2\gamma^5(\sigma^1)^i{}_j\chi^j=0}\, ,\label{eq:Killingspinoreqtsphi1}
\\[4pt]
 &\nonumber\medmath{2\partial_z \chi^i+\frac{\sqrt{f}}{2}\gamma_{31}\chi^i+\frac{Q}{r^2}\gamma_{31}\frac{r}{\ell}\gamma_3(i\sigma^2)^i{}_j\chi^j-i\frac{r}{\ell^2}\gamma_3\gamma^5(\sigma^1)^i{}_j\chi^j=0}\,. \,\\\label{eq:Killingspinoreqtsz1}  
\end{flalign}
Let us assume the $r$-dependence at leading order in the large $r$-limit to be \footnote{The next-to-leading order in the $r$-dependence of the spinors \eqref{eq:Killingspinors} is $\sim r^{-7/2}$, therefore it is safe to solve the Killing spinor equations at leading order.} $\chi^i(r)\sim r^{1/2}$. By further assuming that the spinor should not depend on $(t,z)$, then an expansion of the radial functions \eqref{eq:fr}, \eqref{eq:derf} and
\begin{eqnarray}\label{eq:largerlimitsqrtf}
\sqrt{f}=\frac{r}{\ell}-\frac{\ell\mu}{2r^2}-\frac{\ell Q^2}{2r^3}+\mathcal{O}\left(r^{-4}\right)\,,
\end{eqnarray}
shapes the Killing spinor equation \eqref{eq:Killingspinoreqtsz1} into
\begin{equation}\label{eq:projeqt1}
\frac{r}{\ell^2}\gamma_3\left(\delta^i_j-i\gamma_1\gamma^5(\sigma^1)^i{}_j\right)\chi^j=0\, ,
\end{equation}
which is easily solved by requiring a structure
\begin{equation}\label{eq:structure1}
\chi^i=r^{1/2}\mathbb{P}^i{}_j\mathcal{O}^j{}_k(\phi)\epsilon_0^k\, ,\; \mathbb{P}^i{}_j\equiv\frac{1}{2}\left(\delta^i_j+i\gamma_1\gamma^5(\sigma^1)^i{}_j\right)\, ,
\end{equation}
namely a projector $\mathbb{P}$, a $\phi$-dependent operator carrying  SU(2) indices $\mathcal{O}$, and a doublet of arbitrary constant Majorana spinors $\epsilon_0^i$. Plugging this structure in the large $r$-limit of \eqref{eq:Killingspinoreqtsphi1}, one finds \footnote{In the large $r$-limit, \eqref{eq:Killingspinoreqtsphi1} factorises into two separate equations, \eqref{eq:phieq2} of order $r^{1/2}$, and another one of order $r^{3/2}$, which is nonetheless identically satisfied once given \eqref{eq:structure1}.}
\begin{flalign}\label{eq:phieq2}
\nonumber &\partial_\phi \chi^i +i\frac{Q}{\ell r_0}\gamma^5(\sigma^3)^i{}_j\chi^j=0\, \implies \\[4pt]
&\partial_\phi \mathcal{O}^i{}_j +i\frac{Q}{\ell r_0}\gamma^5(\sigma^3)^i{}_j\mathcal{O}^j{}_k=0\, ,
\end{flalign}
whose solution is
\begin{equation}\label{eq:phiOp2}
\mathcal{O}^i{}_j(\phi)=\text{Exp}\left[-i\frac{Q}{\ell r_0}\gamma^5(\sigma^3) \phi\right]^i{}_j\, ,
\end{equation}
thereby validating \eqref{eq:Majoranaspinorsexplicitform}, \eqref{eq:Majoranaspinorsexplicitformnicer}.
\subsection{Asymptotic superalgebra}
The computation of the anticommutator of the two Noether supercharges, following \eqref{eq:superchargesanticomm}, proceeds as follows. Inserting into the right-hand-side the relevant operator $\tilde{\mathcal{D}}$ and the asymptotic Killing spinors  \eqref{eq:Majoranaspinorsexplicitformnicer}, the former reads
\begin{flalign} \label{eq:Secondanticomm}
\nonumber&\medmath{\{\mathcal{Q},\mathcal{Q}\}=-i\int_{\partial V}d\Sigma_{tr}\varepsilon^{tr\rho\sigma}\left(\bar{\chi}_i\gamma^5\gamma_\rho\tilde{\mathcal{D}}_\sigma\chi^i\right)}\\[2pt]
&\nonumber\medmath{=-2i\int dz d\phi\; \bar{\chi}_i\gamma^5\left(\gamma_\phi \tilde{\mathcal{D}}_z \chi^i-\gamma_z\tilde{\mathcal{D}}_\phi \chi^i\right)}\\[4pt]
\nonumber&\medmath{=-2i\int dz d\phi \;\bar{\chi}_i\gamma^5}\,\times\\
&\nonumber\hspace{0.11cm}\medmath{\left[\sqrt{f}\gamma_2\left(\frac{\omega_z^{ab}}{2}\gamma_{ab}\chi^i+ \frac{Q}{r^2}\gamma^{r\phi}\gamma_z(i\sigma^2)^i{}_j\chi^j-\frac{i}{\ell}\gamma_z \gamma^5(\sigma^1)^i{}_j\chi^j\right)\right.}\\
\nonumber &\hspace{0.3cm} \medmath{\left.\left.-\frac{r}{\ell}\gamma_3\left(2\partial_\phi \chi^i+\frac{\omega_\phi^{ab}}{2}\gamma_{ab}\chi^i-\frac{2Q}{\ell}\left(\frac{1}{r}-\frac{1}{r_0}\right)\gamma^5(i\sigma^3)^i{}_j\chi^j\right.\right.\right.}\\
\nonumber &\hspace{2.6cm}\medmath{\left.\left.+\frac{Q}{r^2}\gamma^{r \phi}\gamma_\phi (i\sigma^2)^i{}_j\chi^j-\frac{i}{\ell}\gamma_\phi \gamma^5(\sigma^1)^i{}_j\chi^j\right)\right]}\\[6pt]
%
% \nonumber &=&-i\int dzd\phi \,\bar{\chi}_i\gamma^5\left[\left(\frac{f}{\ell}\gamma_{231}\chi^i+\frac{Q\sqrt{f}}{r\ell}\gamma_2\gamma^{12}\gamma_3(i\sigma^2)^i{}_j\chi^j-\frac{ir\sqrt{f}}{\ell^2}\gamma_{23}\gamma^5(\sigma^1)^i{}_j\chi^j\right)\right.\\
% %
% \nonumber&& \hspace{3cm}\left. -2\frac{r}{\ell}\gamma_3\partial_\phi \chi^i-\frac{r f^\prime}{2\ell}\gamma_{321}\chi^i+\frac{2Q}{\ell^2}\left(1-\frac{r}{r_0}\right)\gamma_3 \gamma^5(i\sigma^3)^i{}_j\chi^j\right.\\
% %
% \nonumber && \hspace{4.5cm}\left.-\frac{Q\sqrt{f}}{r\ell}\gamma_3\gamma^{12}\gamma_{2}(i\sigma^2)^i{}_j\chi^j+\frac{ir\sqrt{f}}{\ell^2}\gamma_{32}\gamma^5(\sigma^1)^i{}_j\chi^j\right]\\[5pt]
%
\nonumber &\medmath{=-2i\int dzd\phi\;\bar{\chi}_i\gamma^5} \times\\
\nonumber&\hspace{1.1cm}\medmath{\left[\left(-\frac{2r}{\ell}\gamma_3\partial_\phi \chi^i+\frac{2Q}{\ell^2}\left(1-\frac{r}{r_0}\right)\gamma_3\gamma^5(i\sigma^3)^i{}_j\chi^j\right)\right.}\\
\nonumber &\hspace{1.9cm}\medmath{\left.+\left(\frac{f}{\ell}+\frac{rf^\prime}{2\ell}\right)\gamma_{231}\chi^i-2i\frac{r\sqrt{f}}{\ell^2}\gamma_{23}\gamma^5(\sigma^1)^i{}_j\chi^j\right]\, }.\\[2pt]
\end{flalign}
By plugging in the additional $r$-dependence of the spinors $\chi^i\sim r^{1/2}\tilde{\chi}^i$, and taking into account the useful gamma matrix identities
\begin{flalign}\label{eq:gammacontractions}
&\nonumber\gamma^5\gamma_3=-i\gamma^{012}\, ,\quad \gamma^5 \gamma_{231}=i\gamma^0\, ,\\
&\gamma^5\gamma_{23}\gamma^5=\gamma_{23}\,, \quad \gamma^5 \gamma_3\gamma^5=-\gamma_3\, ,
\end{flalign}
equation \eqref{eq:Secondanticomm} is rewritten as
\begin{flalign}\label{eq:anticommforMajorana}
\nonumber&\medmath{\{\mathcal{Q},\mathcal{Q}\}=2\int dz d\phi \,\bar{\tilde{\chi}}_i\; \times}\\
\nonumber & \hspace{1.5cm}\medmath{\left[\frac{2r^2}{\ell}\gamma^{012}\partial_\phi \tilde{\chi}^i-\frac{2Q}{\ell^2}\left(r-\frac{r^2}{r_0}\right)\gamma_3(\sigma^3)^i{}_j\tilde{\chi}^j\right.}\\[3pt]
\nonumber &\hspace{2.1cm}\medmath{\left.+\left(\frac{rf}{\ell}+\frac{r^2f^\prime}{2\ell}\right)\gamma^{0}\tilde{\chi}^i-2\frac{r^2\sqrt{f}}{\ell^2}\gamma_{23}(\sigma^1)^i{}_j\tilde{\chi}^j\right]\,} .\\
\end{flalign}
In the large $r$ limit, \eqref{eq:fr}, \eqref{eq:derf} and \eqref{eq:largerlimitsqrtf} yield
\begin{flalign}\label{eq:largerlimitMajo}
\nonumber&\medmath{\{\mathcal{Q},\mathcal{Q}\}=2\int dz d\phi \,\bar{\tilde{\chi}}_i\;\times}\\
\nonumber & \hspace{1.5cm}\medmath{\left[\frac{2r^2}{\ell}\gamma^{012}\partial_\phi \tilde{\chi}^i-\frac{2Q}{\ell^2}\left(r-\frac{r^2}{r_0}\right)\gamma_{3}(\sigma^3)^i{}_j\tilde{\chi}^j\right.}\\[3pt]
\nonumber&\hspace{1.7cm}\medmath{\left.+\left( \frac{2r^3}{\ell^3}-\frac{\mu}{2\ell}\right)\gamma^0 \tilde{\chi}^i-\left(\frac{2r^3}{\ell^3}-\frac{\mu}{\ell}\right)\gamma_{23}(\sigma^1)^i{}_j\tilde{\chi}^j\right]\,} .\\
\end{flalign}
By using the definition $\bar{\tilde{\chi}}^i=\overline{(\mathbb{P}\mathcal{O})^i{}_j \epsilon_0^j}\equiv ((\mathbb{P}\mathcal{O})^i{}_j \epsilon_0^j)^\dagger\gamma^0=(\epsilon_0^j)^T (\mathcal{O}^k{}_j)^\dagger (\mathbb{P}^i{}_k)^\dagger\gamma^0$ of the Dirac conjugate, one recovers the expression
\begin{flalign}\label{eq:largerlimitMajo2}
\nonumber&\medmath{\{\mathcal{Q},\mathcal{Q}\}=2\int dz d\phi \,(\epsilon_0^j)^T (\mathcal{O}^l{}_j)^\dagger (\mathbb{P}^i{}_l)^\dagger}\\
\nonumber&\hspace{1.5cm}\medmath{\left[-\frac{2r^2}{\ell}\gamma^{12}\partial_\phi \tilde{\chi}^i-\frac{2Q}{\ell^2}\left(r-\frac{r^2}{r_0}\right)\gamma^0\gamma_{3}(\sigma^3)^i{}_j\tilde{\chi}^j\right.} \\
&\nonumber\hspace{1.6cm}\medmath{\left.-\left( \frac{2r^3}{\ell^3}-\frac{\mu}{2\ell}\right) \tilde{\chi}^i-\left(\frac{2r^3}{\ell^3}-\frac{\mu}{\ell}\right)\gamma^0\gamma_{23}(\sigma^1)^i{}_j\tilde{\chi}^j\right]\, }.\\
\end{flalign}
\eqref{eq:largerlimitMajo2} can now be analysed analytically. Inspection of the third line tells us, after some Clifford algebra, that
\begin{align}
&\nonumber\medmath{(\mathcal{O}^l{}_j)^\dagger (\mathbb{P}^i{}_l)^\dagger\;\times}\\
\nonumber & \hspace{0.5cm}\medmath{\left(-\left( \frac{2r^3}{\ell^3}-\frac{\mu}{2\ell}\right) \tilde{\chi}^i-\left(\frac{2r^3}{\ell^3}-\frac{\mu}{\ell}\right)\gamma^0\gamma_{23}(\sigma^1)^i{}_j\tilde{\chi}^j\right)}\\[3pt]
&\nonumber\hspace{0.7cm}\medmath{=\left(-\frac{\mu}{2 \ell}\right)(\mathcal{O}^l{}_j)^\dagger (\mathbb{P}^i{}_l)^\dagger\mathbb{P}^i{}_k\mathcal{O}^k{}_m\epsilon_0^m=\left(-\frac{\mu}{2 \ell}\right)\mathbb{P}^i{}_m\epsilon_0^m\,},\\[1pt]
\end{align}
where we have used the identity 
\begin{equation}(\mathcal{O}^l{}_j)^\dagger (\mathbb{P}^i{}_l)^\dagger\mathbb{P}^i{}_k\mathcal{O}^k{}_m=(\mathcal{O}^l{}_j)^\dagger \mathbb{P}^l{}_k\mathcal{O}^k{}_m=\mathbb{P}^j{}_m\, .
\end{equation} 
As we see, the result is a finite term. The second line instead can be rearranged as
\begin{flalign}\label{eq:cumbersometerm}
&\nonumber\medmath{(\mathcal{O}^l{}_j)^\dagger (\mathbb{P}^i{}_l)^\dagger\left(-\frac{2r^2}{\ell}\gamma^{12}\partial_\phi \tilde{\chi}^i-\frac{2Q}{\ell^2}\left(r-\frac{r^2}{r_0}\right)\gamma^0\gamma_{3}(\sigma^3)^i{}_j\tilde{\chi}^j\right)}\\[3pt]
% \nonumber&=&\frac{2r^2}{\ell}(\mathcal{O}^l{}_j)^\dagger (\mathbb{P}^i{}_l)^\dagger\gamma^{12}\left[-\partial_\phi \tilde{\chi}^i-\frac{Q}{\ell}\left(\frac{1}{r}-\frac{1}{r_0}\right)\gamma^{21}\gamma^0\gamma_{3}(\sigma^3)^i{}_j\tilde{\chi}^j\right]\\[3pt]
% \nonumber &=&\frac{2r^2}{\ell}(\mathcal{O}^l{}_j)^\dagger (\mathbb{P}^i{}_l)^\dagger\gamma^{12}\left[-\partial_\phi \mathbb{P}^i{}_k\mathcal{O}^k{}_m\epsilon_0^m+i\frac{Q}{\ell}\left(\frac{1}{r}-\frac{1}{r_0}\right)\gamma^5(\sigma^3)^i{}_j \mathbb{P}^j{}_k\mathcal{O}^k{}_m\epsilon_0^m\right]\\[3pt]
% \nonumber &=&\frac{2r^2}{\ell}(\mathcal{O}^l{}_j)^\dagger (\mathbb{P}^i{}_l)^\dagger\gamma^{12}\left[-\partial_\phi \mathbb{P}^i{}_k\mathcal{O}^k{}_m\epsilon_0^m+i\frac{Q}{\ell}\left(\frac{1}{r}-\frac{1}{r_0}\right)\mathbb{P}^i{}_k\gamma^5(\sigma^3)^k{}_j \mathcal{O}^j{}_m\epsilon_0^m\right]\\[3pt]
\nonumber &\hspace{1.5cm}\medmath{= \frac{2r^2}{\ell}(\mathcal{O}^l{}_j)^\dagger (\mathbb{P}^i{}_l)^\dagger\gamma^{12}\mathbb{P}^i{}_k}\times\\[1pt]
&\nonumber\hspace{1.8cm}\medmath{\left[-\partial_\phi \mathcal{O}^k{}_m+i\frac{Q}{\ell}\left(\frac{1}{r}-\frac{1}{r_0}\right)\gamma^5(\sigma^3)^k{}_j \mathcal{O}^j{}_m\right]\epsilon_0^m\,} .\\
\end{flalign}
Due to the very structure of the asymptotic Killing spinors, this term is zero. Indeed, it is not hard to prove that 
\begin{equation}
(\mathbb{P}^i{}_l)^\dagger\gamma^{12}\mathbb{P}^i{}_k=\mathbb{P}^l{}_i \gamma^{12}\mathbb{P}^i{}_k=\gamma^{12}\bar{\mathbb{P}}^l{}_i \mathbb{P}^i{}_k=0\, ,
\end{equation}
where $\bar{\mathbb{P}}^l{}_i\equiv \tfrac{1}{2}\left(\delta^i_k-i \gamma_1\gamma^5 (\sigma^1)^i{}_k\right)$.

In order to read off the superalgebra, therefore, we can focus only on the third line of \eqref{eq:largerlimitMajo2}, stripping out of the anticommutator the constant Majorana spinors. The final expression,
\begin{flalign}\label{eq:superalgebra1}
\nonumber&\epsilon_{0\,\alpha i}\{\mathcal{Q}^{\alpha i},\mathcal{Q}_{\beta j}\}\epsilon_0^{\beta j}=\epsilon_{0\,\alpha i}\int dz \left(-\frac{\Delta  \mu}{ \ell}\right)\mathbb{P}^{\alpha\, i}{}_{\beta\, j}\epsilon_0^{\beta\, j}\\[3pt]
&=\epsilon_{0\,\alpha i}\int dz \left(-\frac{\Delta  \mu}{ \ell}\right)\left(\delta^\alpha_\beta \delta^i_j+i(\gamma_{1}\gamma^5)^\alpha{}_\beta (\sigma^1)^i{}_j\right)\epsilon_0^{\beta\, j}\, ,
\end{flalign}
is obtained, where the integral on $\phi$ has been performed.
\subsection{BPS bound}
Being both $\mathcal{Q}_1$ and $\mathcal{Q}_2$ reals, let us define the complex operators
\begin{eqnarray}\label{eq:complexcharges1}
\nonumber\mathbf{Q}^\alpha{}_\beta&=&\frac{1}{\sqrt{2}}\left(\mathcal{Q}_1{}^\alpha{}_\beta+i\mathcal{Q}_2{}^\alpha{}_\beta\right)\, ,\\[1pt] \overline{\mathbf{Q}}^\alpha{}_\beta&=&\frac{1}{\sqrt{2}}\left(\mathcal{Q}_1{}^\alpha{}_\beta-i\mathcal{Q}_2{}^\alpha{}_\beta\right)\, ,
\end{eqnarray}
and compute
\begin{equation}\label{eq:complexanticomm1}
\{\mathbf{Q},\overline{\mathbf{Q}}\}^\alpha{}_\beta=\int dz\left(\frac{-\Delta  \mu}{ \ell} \right)\delta^\alpha_\beta\, .
\end{equation}
The structure of this anticommutator entails the product of a complex number and of its complex conjugate, and must therefore be semi-positive definite. This leads to the BPS bound
\begin{equation}\label{eq:BPS1}
\text{Tr}\left[\{\mathbf{Q},\overline{\mathbf{Q}}\}\right]\equiv\{\mathbf{Q},\overline{\mathbf{Q}}\}^\alpha{}_\alpha \geq 0 \quad \Longrightarrow \boxed{ \quad \mu \leq 0 \;\;\;}\, ,
\end{equation}
which, as expected, imposes positivity ($\mu$ corresponds to minus the energy density, see \eqref{eq:energymomtensor}) of energy for all physical solutions, and it is saturated by the supersymmetric soliton.
\section*{Acknowledgements}
The research of AA is supported in part by the Fondecyt Grants 1210635, 1221504, 1230853 and 1200986 and by the FAPESP of Sao Paulo with a Visiting Researcher Award, 2022/11765-7. The work of MC is supported by the Spanish Agencia Estatal de Investigacion through the grant “IFT Centro de Excelencia Severo Ochoa SOLAUT$\_$00044994”, and by the grant PID2021-123017NB-I00, funded by MCIN/AEI/10.13039/50110001103. MC also gratefully acknowledges la Caixa Foundation (ID100010434), that partially supported the early stages of this work and financed a mid-term stay at the Politencico di Torino.
\bibliography{references2} 
\end{multicols}
\end{document}